\documentclass[a4paper, 10pt]{article}
\usepackage{amsmath,amssymb,bm}
\usepackage[dvipdfm]{graphicx}
\usepackage{color}
\usepackage{ulem}
\usepackage{mathrsfs} 
\title{Influence of synaptic depression on memory storage capacity}
\author{Yosuke Otsubo$^{1}$,Kenji Nagata$^{1}$, Masafumi Oizumi$^{2,3}$, Masato Okada$^{1,3\dagger}$
}

\begin{document}
\maketitle
\section*{Abstract}
Synaptic efficacy between neurons is known to change within a short time scale dynamically. 
Neurophysiological experiments show that high-frequency presynaptic inputs decrease synaptic efficacy between neurons.
This phenomenon is called synaptic depression, a short term synaptic plasticity. 
Many researchers have investigated how the synaptic depression affects the memory storage capacity. 
However, the noise has not been taken into consideration in their analysis.
By introducing ``temperature'', which controls the level of the noise, into an update rule of neurons, 
we investigate the effects of synaptic depression on the memory storage capacity in the presence of the noise. 
We analytically compute the storage capacity by using a statistical mechanics technique 
called Self Consistent Signal to Noise Analysis (SCSNA). 
We find that the synaptic depression decreases the storage capacity in the case of finite temperature 
in contrast to the case of the low temperature limit, where the storage capacity does not change. 
\footnote{ {\it Graduate School of Frontier Sciences, The University of Tokyo, Kashiwa, Chiba 277-8561}\\
$^2$ {\it Research Fellow of the Japan Society for the Promotion of Science}\\
$^3$ {\it Brain Science Institute, RIKEN, Wako, Saitama, 351-0198}\\
$\dagger$E-mail address: okada@k.u-tokyo.ac.jp\\
}
\clearpage
\section{Introduction}

Synaptic efficacy between neurons is known to change within a short time scale dynamically. 
Neurophysiological experiments show that high-frequency presynaptic inputs decrease synaptic efficacy between neurons.
This phenomenon is called synaptic depression, a short term synaptic plasticity. \cite{Thomson}\cite{Abbott}\cite{Tsodyks}
Many researchers have investigated how the synaptic depression affects the retrieval property of 
the associative memory model.\cite{Pantic}\cite{Otsubo}
The associative memory model is a typical artificial neural network model that has discretely 
distributed fixed-point attractors as stored patterns. \cite{Hopfield}
Otsubo \textit{et al.} showed that synaptic depression destabilizes a steady state, and then the network 
oscillates among attractors analytically. \cite{Otsubo}
Although their research focused on network stored finite memory patterns, we should focus on the infinite version
in order to discuss the general retrieval property of network. In the case that synaptic strength is static, 
the associative memory model has the maximum number of memory patterns that the network is able to retrieve. 
The number of such patterns per neuron is called storage capacity. 
This is important quantity for discussing memory performance\cite{Fusi}. 
It is interesting problem how short term synaptic plasticity affects the stability of network with infinite attractors. 
Therefore, we focus on the storage capacity in order to discuss the retrieval property of the network with synaptic depression 
in the case of infinite memory patterns.
There have been previous investigations into how the synaptic depression influences the storage capacity.\cite{Torres}\cite{Matsumoto}\cite{Mejias} 
Matsumoto \textit{et al.} showed that the synaptic depression does not affect the storage capacity in the low temperature limit.\cite{Matsumoto} 
However, the noise has not been taken into consideration in their analysis. 
The storage capacity is known to depend on ``temperature", which controls the level of the noise.\cite{Amit}

In this paper,  
we investigate the effects of synaptic depression on the memory storage capacity in the presence of temperature.
We analytically compute the storage capacity by using a statistical mechanics technique 
called Self Consistent Signal to Noise Analysis (SCSNA), which is an extended method of signal to noise analysis\cite{Shiino}. 
This allows us to discuss the retrieval property in the network with synaptic depression more generally.\par  
This paper consists of five sections. 
Section $2$ describes the model used in this paper. 
Section 3 introduces the analysis by SCSNA in order to obtain 
order parameter equations. 
Section 4 investigates the steady state of network with synaptic depression by 
theory and computer simulation. Section 5 summarizes the results obtained in this paper discusses them.

\section{Model}
We consider a recurrent neural network model with $N$ fully connected analogue neurons. 
The state of $i$-th neuron $m_i(t)$, which has a continuous value between $0$ and $1$, i.e., $[0,1]$, 
and an internal potential $h_i(t)$ at time $t$ 
are updated by the following rule\cite{York}:
\begin{eqnarray}
\label{eq:neuron_m}
\label{eq:m_input}
\lefteqn{} &&m_i(t+1)=F(h_i(t)),\\
&& h_i(t)=  \sum_{j\neq i}J_{ij}(t)m_j(t),
\end{eqnarray}
where $J_{ij}(t)$ represents the synaptic weight between $i$-th neuron 
and $j$-th neuron at time $t$, and
$F(\cdot)$ is an output function. 
Thus, the state of network at time $t$ is 
characterized by $\textrm{\boldmath $m$}(t)=(m_1(t),\cdots,m_N(t))$. 
The synaptic weight $J_{ij}(t)$ incorporating synaptic depression can be described by
\begin{eqnarray}
\lefteqn{}J_{ij}(t)&=&\tilde{J}_{ij} x_j(t),\\
x_j(t+1)&=&x_j(t)+\frac{1-x_j(t)}{\tau}-U_{_{SE}}x_j(t)m_j(t),
\label{eq:x_dym}
\end{eqnarray}
where $x_j(t)$ is determined by a phenomenological model of synapses,\cite{Abbott,Tsodyks,Pantic} 
and takes $0<x_j(t)\leq 1$, where $x_j(t)=1$ 
corresponds to the case without synaptic depression. 
The depression variable $x_j(t+1)$ from the presynaptic neuron is decreased 
by a certain fraction $U_{SE}x_j(t)$ 
and recovers with time constant $\tau$. 
The fixed synaptic weight $\tilde{J_{ij}}$ is, according to the Hebbian rule, set to
\begin{equation}
\label{eq:hebb}
\tilde{J}_{ij}=\frac{1}{N}\sum_{\mu=1}^{\alpha N }\xi_{i}^{\mu}\xi_{j}^{\mu}, \\
\end{equation}
where $\alpha N$ corresponds to the number of memory patterns, 
and $\alpha$ is called the loading rate, which is
the ratio between the number of embedded memory patterns and neurons. 
Each element of the memory pattern 
$\textrm{\boldmath $\xi$}^{\mu}=(\xi_1^{\mu},\cdots \xi_N^{\mu})$ is generated randomly and independently by 
\begin{equation}
\mathrm{Prob}[\xi^{\mu}_i=\pm 1]=\frac{1}{2}.
\label{eq:prob_xi}
\end{equation}

\section{Analysis of Steady State by SCSNA}
In this section, we derive macroscopic parameters in steady state 
in the case of $\alpha \sim O(1)$ by using SCSNA\cite{Shiino}, which is an extension of a naive signal to noise (S/N) analysis. \par
Considering the steady states, $m_i(\infty)\equiv m_i$, $h_i(\infty)\equiv h_i$ 
and $x_i(\infty)\equiv x_i$,
we obtain the following equations from eqs. (\ref{eq:m_input}) and (\ref{eq:x_dym}), 
\begin{eqnarray}
\label{eq:m_steady}
\lefteqn{}m_i&=&F\left(h_i\right), \\
\label{eq:input_steady}
h_i&=&\sum_{j \neq i}\tilde{J}_{ij}x_j m_j,\\
 x_j&= &\frac{1}{1+\gamma m_j} \:\:(\gamma \equiv U_{_{SE}}\tau),
\end{eqnarray}
where $\gamma$ indicates the level of synaptic depression in the steady state. 
Here, we use the following transformation, 
\begin{equation}
r_j\equiv x_jm_j=\frac{m_j}{1+\gamma m_j},
\label{eq:r_m}
\end{equation}
and then we obtain the following relationship concerning $r_j$ and internal 
potential $h_j$ for eq. (\ref{eq:m_steady}) and (\ref{eq:input_steady}),
\begin{eqnarray}
\label{eq:r_F}
r_j&=&\frac{F(h_j)}{1+\gamma F(h_j)}\equiv G(h_j),\\
h_j&\equiv& \sum_{j\neq i}\tilde{J}_{ij}r_j.
\label{eq:h_j0}
\end{eqnarray}
We introduce the following order parameter, 
\begin{eqnarray}
\pi^{\mu}_r&\equiv& \frac{2(1+\gamma)}{N}\sum_{i=1}^{N}\xi_i^{\mu}r_i,
\label{eq:r_overlap_def}
\end{eqnarray}
where $\pi^{\mu}_r$ corresponds to overlap between $\textrm{\boldmath $r$}$ and $\textrm{\boldmath $\xi$}^{\mu}$.
If the state of network retrieve a memory pattern $\textrm{\boldmath $\xi$}^{\mu}$, 
the above order parameter is exactly $1$ at $N\rightarrow \infty$, i.e., $\pi_r^{\mu}=1$.
By using $r_j$ in place of $m_j$, the effect of synaptic depression 
is incorporated in the function $G(\cdot)$. 
Thus, we can apply SCSNA to our model under the transformation (\ref{eq:r_m}).
The basis of the SCSNA is in the systematic splitting of the internal potential into a signal and 
a cross-talk noise. Moreover, the cross-talk noise part consists of two elements. One is an effective 
self-coupling term that comes from statistical correlations caused by the recurrent connections, and 
the other obeys Gaussian distribution. 
By this scheme, the order parameter equations are obtained as follows:
\begin{eqnarray}
\label{eq:order_1}
Y&=&G\left(\frac{\xi \pi_r}{2(1+\gamma)}+\sigma \tilde{z}+\Gamma Y\right),\: (\tilde{z}\sim \mathscr{N}(0,1)).\\
 \pi_r&= &2(1+\gamma)\int Dz \langle\langle \xi Y \rangle\rangle  .\\
 q&= &\int Dz \langle\langle Y^2 \rangle\rangle .\\
 U& =&\frac{1}{\sigma}\int Dz z\langle\langle Y \rangle\rangle .\\
 \Gamma& =&\frac{\alpha U}{1-U} .\\
 \sigma &=& \frac{\alpha q}{(1-U)^2}.
 \label{eq:order_2}
\end{eqnarray}
The variable $Y$ represents the effective output for $r_i$ and the stochastic variable $\xi$, 
obeying eq. (\ref{eq:prob_xi}), corresponds to a retrieving pattern component $\textrm{\boldmath $\xi$}^1$.
The order parameters $\pi_r$ correspond to overlap $\pi_r^1$. 
Note that the bracket $\langle\langle \cdot \rangle\rangle$ mean the expectation for stochastic variable $\xi$ 
and $Dz=1/\sqrt{2\pi}\exp(-z^2/2)$.
The notation $\mathscr{N}(0,1)$ represents Gaussian distribution with 0 mean 1 variance. 
Each argument of the function $G(\cdot)$  consists of three terms. 
The first term, $\frac{\xi \pi_r}{2(1+\gamma)}$, comes from a signal component, the second term is 
assumed to be Gaussian distribution with 0 mean $\sigma^2$ variance, and 
the third term means systematic bias of crosstalk noise. 
The detailed derivation of the above equations is given in the appendix. \par
We should also consider the overlap between the state of network, $\textrm{\boldmath $m$}$, and 
$\mu-th$ memory pattern, $\textrm{\boldmath $\xi$}^{\mu}$, in order to describe the macroscopic state of network at a steady state, 
\begin{eqnarray}
\label{eq:overlap_0}
\pi_{m}^{\mu}\equiv \frac{1}{N}\sum_{i=1}^{N}\xi_i^{\mu}(2m_i-1).
\end{eqnarray}
By using $r_i$, the overlap can be described as
\begin{eqnarray}
\label{eq:pi_m}
\pi_m= \int Dz \left\langle\left\langle \xi \frac{(2+\gamma)Y-1}{1-\gamma Y} \right\rangle\right\rangle,
\end{eqnarray}
where $\pi_m\equiv \pi_m^1$.\par
In this paper we consider the output function as sigmoidal,
\begin{equation}
F(h)=\frac{1}{2}(1+\tanh \beta h).
\label{eq:sigmoid}
\end{equation}
Here, $\beta\equiv \frac{1}{T}$ represents the inverse temperature, and the neuron state, 
$m_i$, takes a binary value $1$ or $0$ in $\beta\rightarrow \infty$. 
The previous work estimated the storage capacity 
under the following relationship at the low temperature limit of $\beta\rightarrow \infty$\cite{Matsumoto}:
\begin{eqnarray}
\label{eq:Matsumoto_ap}
x_jm_j&=&\frac{m_j}{1+\gamma m_j}=\frac{m_j}{1+\gamma},\\
F(h)&=& \Theta(h),
\label{eq:Matsumoto_ap2}
\end{eqnarray}
where $\Theta(h)$ is a step function:
\begin{eqnarray}
\Theta(h)=\left\{
\begin{array}{ll}
1 \:(h\geq 0) &\quad  \\
0 \: (h<0)&\quad
\end{array}
\right..
\end{eqnarray} 
Therefore, we obtain the relationship, $m_i=\Theta(\sum_{j\neq i} \tilde{J}_{ij} m_j/(1+\gamma))=\Theta(\sum_{j\neq i} \tilde{J}_{ij} m_j)$ 
by using  eqs. (10) and (12). 
This equation is same as one without synaptic depression. 
That means the storage capacity does not change in $\beta\rightarrow \infty$.
In the case of finite $\beta$, the convergent point and the slope of $G(h)$ is different from those of $F(h)$ (Fig. 1). 
For these reasons, the storage capacity is expected to change at finite temperature. 

\begin{figure}[t]
\begin{center}
\includegraphics[width=2.5in]{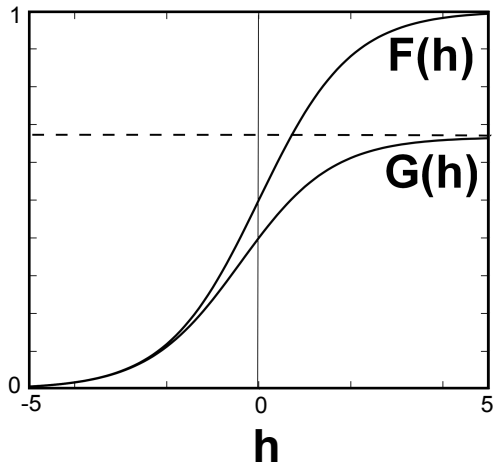}
\caption{
The form of $F(h)$ and $G(h)$ 
in the case of the sigmoidal output function (22) with $\beta=2.0$ and $\gamma=0.5$ as an example.
Dashed line in left figure represents value of $G(h)$ in limit of large $h$ for eq. (\ref{eq:r_F}), i.e., $1/(1+\gamma)$.
}
\label{fig:output}
\end{center}
\end{figure}

\subsection{Results}
In this section, we investigate the storage capacity in a steady state 
for finite temperature 
by using theory and computer simulation numerically. 
We use the sigmoid function, $F(\cdot)=(1+\tanh \beta (\cdot))/2$, for the output function. 
Using sigmoid function make us to find several solution candidates in eq.(\ref{eq:order_1}), 
so that we should introduce the Maxwell rule for choosing a unique solution\cite{Shiino}.
Computer simulations are performed with $N=5000$ and random patterns $\left\{\textrm{\boldmath $\xi$}^{\mu}\right\}$ 
following eq. (\ref{eq:prob_xi}) in 11 trials. 
The initial state in computer simulations is set to $m_i(0)=(\xi_i^1+1)/2$ and $x_i(0)=1$, i.e., $\pi_m^{1}=1,\pi_r^{1}=1$.

\begin{figure}[h]
\begin{center}
\includegraphics[width=3.2in]{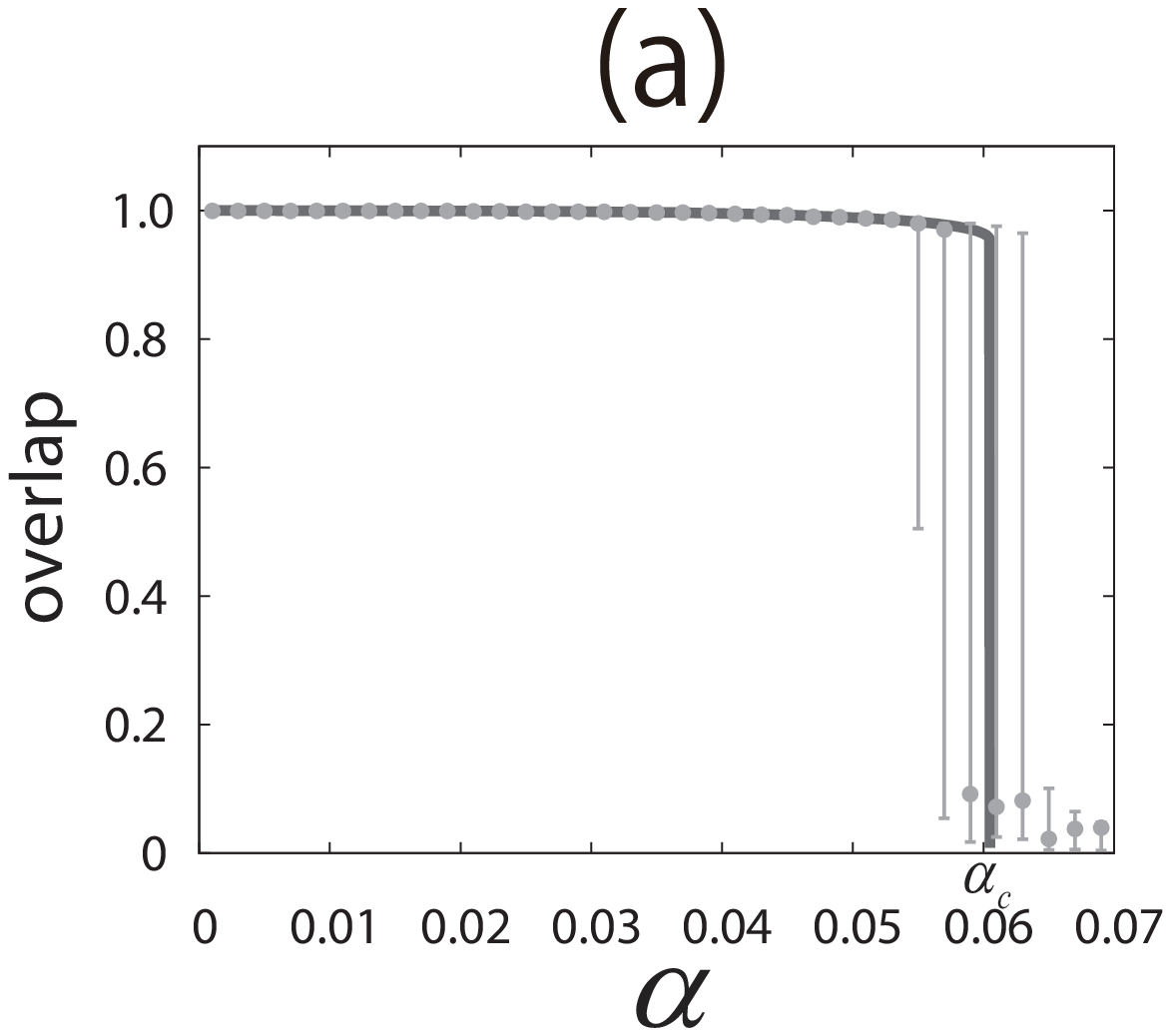}
\includegraphics[width=3.2in]{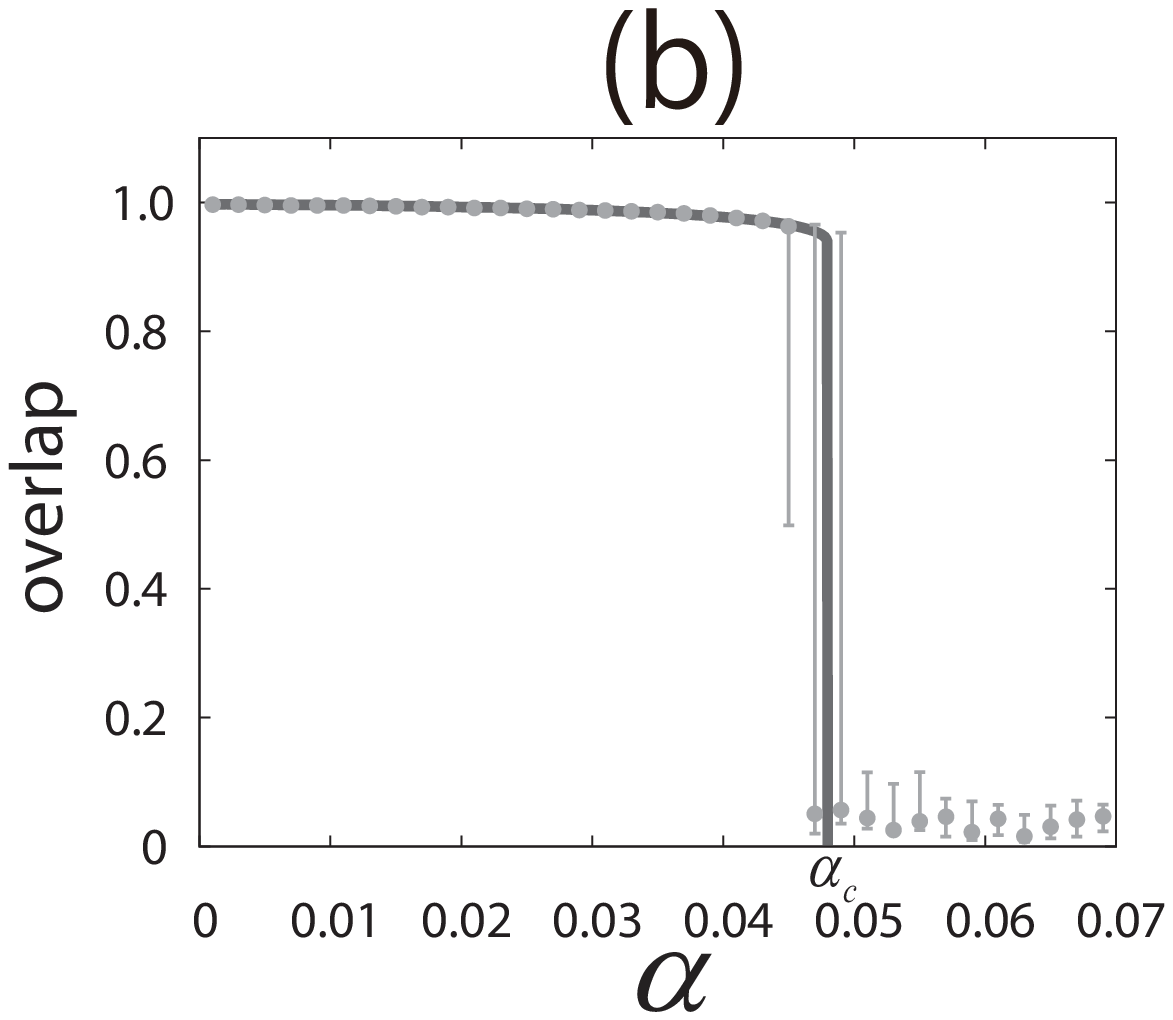}
\caption{
Dependency of $\pi_m$ on $\alpha$ at $T=0.1$. 
Solid line indicates theoretical results by solving (\ref{eq:order_1})-(\ref{eq:order_2}) and (\ref{eq:pi_m}) numerically. 
Error bars indicate medians with minimum and maximum values obtained by computer simulation in $11$ trials at $N=5000$.
(a): case without synaptic depression, i.e., $\gamma=0.0$. (b): case with synaptic depression, i.e., $\gamma=0.5,\tau=2.0,U_{_{SE}}=0.25$.
}
\label{fig:result-1}
\end{center}
\end{figure}
Figure \ref{fig:result-1} (a) and (b) show how the overlap $\pi_m^{1}$ defined by eq. (\ref{eq:overlap_0}) depends on 
the loading rate $\alpha$ at $T=0.1$. 
Figure \ref{fig:result-1} (a) and (b) are the cases without and with synaptic depression, 
i.e., $\gamma=0.5,\tau=2.0$. 
The solid lines show the solutions obtained by solving eqs. (\ref{eq:order_1})-(\ref{eq:order_2}) and (\ref{eq:pi_m})  numerically 
while error bars indicate medians with minimum and maximum values obtained by computer simulation in $11$ trials at $N=5000$. 
The overlap of 1 means the system retrieve $1$-st stored pattern perfectly. 
The storage capacity $\alpha_c$ is the loading rate at $\pi_m^1\rightarrow 0$ in the solid line. \par
 Theoretical results obtained by SCSNA and computer simulation results agree well
because the overlap given by computer simulation drastically decreases at $\alpha_c$.\par
We can see that the storage capacity in the case without synaptic depression is $0.060$ in Fig. \ref{fig:result-1} (a), 
while in the case with synaptic depression it is $0.048$ in Fig. \ref{fig:result-1} (b). 
Therefore, the synaptic depression decreases the storage capacity for finite temperature.\par

\begin{figure}[t]
\begin{center}
\includegraphics[width=3.6in]{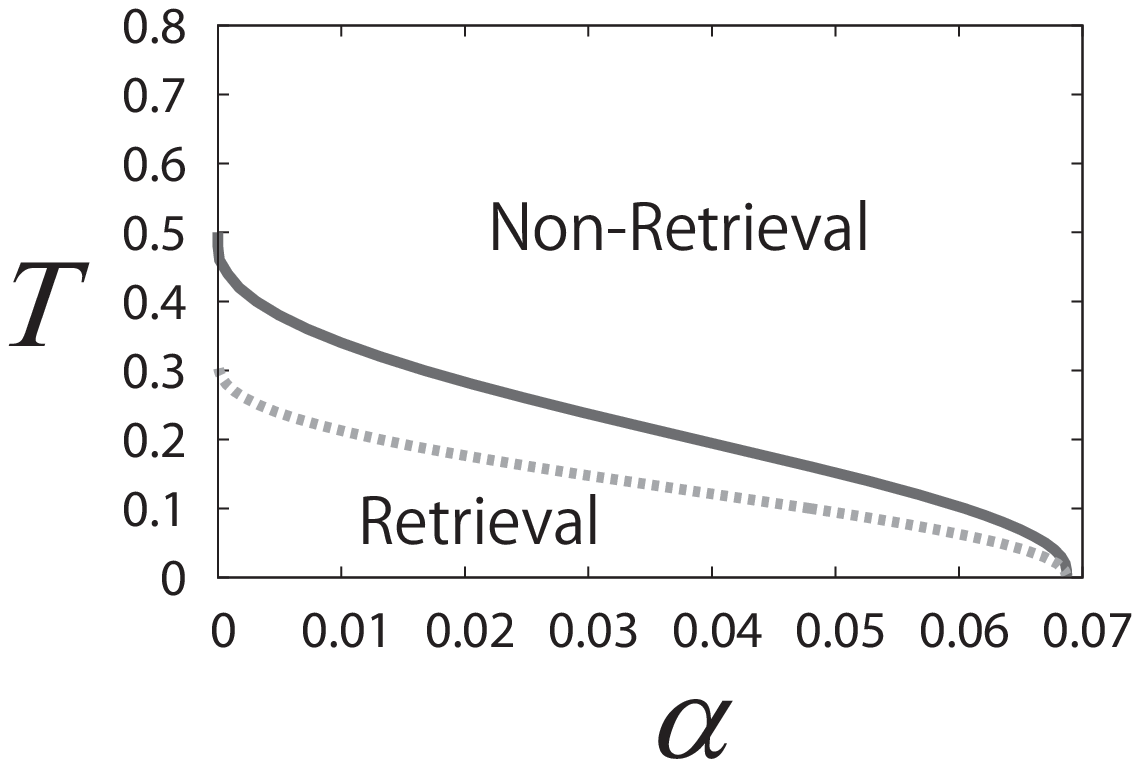}
\caption{Phase diagram representing temperature dependence of storage capacity. 
In retrieval phase, network can retrieve memory pattern, $\pi_m \sim 1$, 
while in non-retrieval phase, network cannot retrieve any memory pattern, $\pi_m \sim 0$. 
Solid line indicates $\alpha_c$ in case without synaptic depression, i.e., $\gamma=0.0$
Dashed line indicates $\alpha_c$ in case with synaptic depression, i.e., $\gamma=0.5$.
}
\label{fig:result-2}
\end{center}
\end{figure}

Next, Fig. \ref{fig:result-2} shows the phase diagram with respect to temperature $T$. 
In this phase diagram, the region in which the network succeeds 
in retrieving $1-$st memory pattern which is the initial state
is denoted as the ``Retrieval" phase, i.e., $\pi_m^1 \sim 1$. 
In the ``Non-Retrieval" phase, the retrieval is not successful i.e., $\pi_m\sim 0$. 
The solid and dashed lines respectively indicate the storage capacity $\alpha_c$ at corresponding temperature by solving 
eqs. (\ref{eq:order_1})-(\ref{eq:order_2}) and (\ref{eq:pi_m}) in the cases without and with
synaptic depression i.e., $\gamma=0.0$, $\gamma=0.5$. 
On the basis of this result, we can see that the synaptic depression does not change the storage capacity in the low temperature limit, 
while the storage capacity decreases for finite temperature. 
These results expand the previous result which argue that synaptic depression does not change 
the storage capacity in the low temperature limit \cite{Matsumoto}. \par
\begin{figure}[t]
\begin{center}
\includegraphics[width=3.5in]{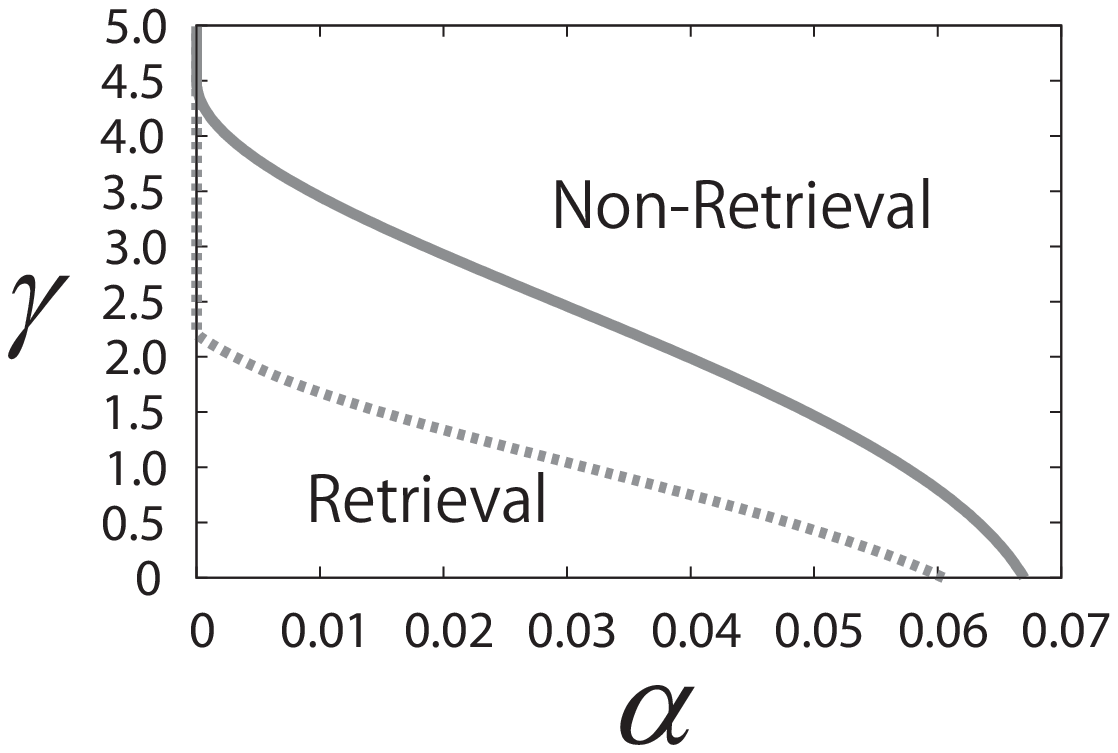}
\caption{
Phase diagram representing level of synaptic depression dependence of storage capacity.
Solid line indicates $\alpha_c$ in case of $T=0.05$. 
Dashed line indicates $\alpha_c$ in 
the case of $T=0.1$.
}
\label{fig:result-2_2}
\end{center}
\end{figure}
Fig. \ref{fig:result-2_2} also shows the diagram with respect to the level of synaptic depression 
at a controlled temperature.  
The solid and the dashed lines indicate the cases of $T=0.05$ and $T=0.1$, respectively. 
These results indicate that the storage capacity decreases when the 
level of synaptic depression increases in finite temperature. 
Furthermore, we can see that the higher the level of synaptic depression, 
the more storage capacity decreases. 

\section{Conclusion}
In this paper, we discussed the associative memory model 
with synaptic depression and  analytically
investigated the storage capacity at finite temperature by SCSNA. 
There have been previous investigations of how the synaptic depression affects the storage capacity\cite{Torres}\cite{Matsumoto}. 
However, their investigations focused on only low temperature limit. 
The analysis introduced in this paper enables us to discuss 
the storage capacity of the model with synaptic depression more generally. 
As a result, we found that the synaptic depression decreased the storage capacity in the case of finite temperature 
in contrast to the result in the case of zero temperature, where synaptic depression does not change the storage capacity. 

\par

\section{Appendix:Detail SCSNA Description}
The goal of this appendix is to derive the order parameter equations 
from (\ref{eq:order_1}) to (\ref{eq:order_2}) by using SCSNA. By using eq. (\ref{eq:r_overlap_def}), we rewrite the internal potential (\ref{eq:h_j0}) as
\begin{eqnarray}
h_i= \frac{1}{2(1+\gamma)}\sum_{\mu=1}^{\alpha N}\xi_i^{\mu}\pi_r^{\mu}-\alpha r_i.
\label{eq:h_ape}
\end{eqnarray}
The output of $r_i$ can be formally expressed by
\begin{eqnarray}
r_i&=&G\left( \frac{1}{2(1+\gamma)}\sum_{\mu=1}^{\alpha N}\xi_i^{\mu}\pi_r^{\mu}-\alpha r_i \right)\\
   &=&\tilde{G}\left(\frac{1}{2(1+\gamma)}\sum_{\mu=1}^{\alpha N}\xi_i^{\mu}\pi_r^{\mu}\right)\\
   &=&\tilde{G}\left(\frac{1}{2(1+\gamma)}\xi_i^{\mu}\pi_r^{\mu}+\frac{1}{2(1+\gamma)}\sum_{\nu\neq \mu}\xi_i^{\nu}\pi_r^{\nu}\right),
   \label{eq:r_G}
\end{eqnarray}
where $\tilde{G}(x)$ represents the solution of $x=G(x)$．
Here, we discuss when the $1$-st pattern $\textrm{\boldmath $\xi$}^{1}$ is retrieved, and then 
the residual overlap $\pi^{\mu}_r \sim O(1/\sqrt{N})\:(\mu\neq 1)$ because $\alpha \sim O(1)$. 
Thus, 
the overlap $\pi_r^{\mu}$ can be denoted as
\begin{eqnarray}
\pi^{\mu}_r&=&\frac{2(1+\gamma)}{N}\sum_{j=1}^N\xi_j^{\mu}r_j\\
           &=&\frac{2(1+\gamma)}{N}\sum_{j=1}^N\xi_j^{\mu}\left(r_j^{(\mu)}+\frac{1}{2(1+\gamma)}\xi_j^{\mu}\pi_r^{\mu}r_j'^{(\mu)}\right)\\
           &=&\frac{2(1+\gamma)}{N}\sum_{j=1}^N\xi_j^{\mu}r_j^{(\mu)}+\frac{1}{N}\pi_r^{\mu}\sum_{j=1}^N r_j'^{(\mu)}\\
           &=&\frac{2(1+\gamma)}{N}\sum_{j=1}^N\xi_j^{\mu}r_j^{(\mu)}+U\pi_r^{\mu},
           \label{eq:pi_r}
\end{eqnarray}
where 
\begin{equation}
U\equiv\frac{1}{N}\sum_{j=1}^{N}r_j'^{(\mu)},
\end{equation}
by using the following approximation of eq. (\ref{eq:r_G}),
\begin{eqnarray}
r_i              &\sim&r_i^{(\mu)}+\frac{1}{2(1+\gamma)}\xi_i^{\mu}\pi_r^{\mu}r_i'^{(\mu)},\\
r_i^{(\mu)} &\equiv&\tilde{G}\left(\frac{1}{2(1+\gamma)}\sum_{\nu\neq \mu}\xi_i^{\nu}\pi_r^{\nu}\right),\\
r_i'^{(\mu)}&\equiv& \tilde{G}'\left(\frac{1}{2(1+\gamma)}\sum_{\nu\neq \mu}\xi_i^{\nu}\pi_r^{\nu}\right),
\end{eqnarray}
where $r_i^{(\mu)}$ is the value drawn the effect of $\mu$-th pattern from $r_i$.
Solving eq. (\ref{eq:pi_r}), the overlap $\pi_r^{\mu}$ can be rewritten as
\begin{equation}
\pi_r^{\mu}=\frac{2(1+\gamma)}{N(1-U)}\sum_{j=1}^N\xi_j^{\mu}r_j^{(\mu)},
\end{equation}
and the internal potential (\ref{eq:h_ape}) is 
\begin{eqnarray}
h_i
   &=&\frac{\xi_i^1\pi_r^{1}}{2(1+\gamma)}+\frac{1}{N(1-U)}\sum_{j=1}^{N}
   \sum_{\mu=2}^{\alpha N}\xi_i^{\mu}\xi_j^{\mu}r_j^{(\mu)}-\alpha r_i\nonumber\\
   &\sim&\frac{\xi_i^1\pi_r^{1}}{2(1+\gamma)}+\frac{1}{N(1-U)}\sum_{j\neq i}^{N}
   \sum_{\mu=2}^{\alpha N}\xi_i^{\mu}\xi_j^{\mu}r_j^{(\mu)}+\frac{\alpha U}{1-U} r_i\nonumber\\
   \label{eq:input_signal_noise}
   &=&\frac{\xi_i^1\pi_r^{1}}{2(1+\gamma)}+\tilde{z}_i+\Gamma r_i,\\
\tilde{z}_i&=&\frac{1}{N(1-U)}\sum_{j\neq i}^{N}\sum_{\mu=2}^{\alpha N}\xi_i^{\mu}\xi_j^{\mu}r_j^{(\mu)},\\
\Gamma&=&\frac{\alpha U}{1-U} r_i.
\end{eqnarray}
where we use the approximation, $r_i^{(\mu)}\sim r_i $, becasue $r_i^{(\mu)}-r_i\sim O(1/\sqrt{N})$. 
The first term in the right-hand side of eq. (\ref{eq:input_signal_noise}) comes from 
signal components, 
the second term is assigned to Gaussian distribution with 0 mean and the following variance,
\begin{eqnarray}
\sigma ^2 &=&E[(z_i)^2]-(E[z_i])^2=\frac{\alpha}{(1-U)^2}q, 
\end{eqnarray}
and the third term means the systematic bias of the crosstalk 
noises. 
Here, we define Edwards-Anderson order parameter as 
\begin{eqnarray}
q&\equiv&\frac{1}{N}\sum_{j\neq i}r_j^{(\mu)2}.
\end{eqnarray}
We obtain the following order parameter equations:
\begin{eqnarray}
Y&=&G\left(\frac{\xi \pi_r}{2(1+\gamma)}+\sigma \tilde{z}+\Gamma Y \right)\: (\tilde{z}\sim \mathscr{N}(0,1)),\\
 \pi_r&= &2(1+\gamma)\int Dz \langle\langle \xi Y \rangle  \rangle,\\
 q&= &\int Dz \langle\langle Y^2 \rangle \rangle ,\\
 U& =&\frac{1}{\sigma}\int Dz z\langle \langle Y \rangle \rangle,\\
 \Gamma& =&\frac{\alpha U}{1-U}, \\
 \sigma &=& \frac{\alpha q}{(1-U)^2}
\end{eqnarray}
where we use the correspondence, 
$\xi_i^{\mu}\rightarrow \xi,\:\pi_r^{\mu}\rightarrow \pi_r,\:r_j \rightarrow Y,\:\pi_{m}^{\mu}\rightarrow \pi_{m}$, $Dz=1/\sqrt{2\pi}\exp(-z^2/2)$, 
and the bracket $\langle \langle \cdot \rangle \rangle$ represents the average with respect to stochastic variable $\textrm{\boldmath $\xi$}^{\mu}$.
The above equations return to the usual order parameter equations in the case without synaptic depression.

\end{document}